\documentclass[pdflatex,sn-mathphys-num]{sn-jnl}

\usepackage{graphicx}%
\usepackage{multirow}%
\usepackage{amsmath,amssymb,amsfonts}%
\usepackage{amsthm}%
\usepackage{mathrsfs}%
\usepackage[title]{appendix}%
\usepackage{xcolor}%
\usepackage{textcomp}%
\usepackage{manyfoot}%
\usepackage{booktabs}%
\usepackage{algorithm}%
\usepackage{algorithmicx}%
\usepackage{algpseudocode}%
\usepackage{listings}%
\usepackage{siunitx}
\usepackage{csquotes}

\theoremstyle{thmstyleone}%
\theoremstyle{thmstyletwo}%

\theoremstyle{thmstylethree}%

\begin{document}

\title[Quantifying correlations between IOL and fake news]{Quantifying correlations between information overload and fake news during COVID-19 pandemic:
a Reddit study with the BERT model approach}


\author[1]{\fnm{Jan} \sur{Rawa}}\email{jan.rawa.dokt@pw.edu.pl}

\author*[1,2]{\fnm{Julian} \sur{Sienkiewicz}}\email{julian.sienkiewicz@pw.edu.pl}

\affil[1]{\orgname{Warsaw University of Technology}, \city{Warsaw}, \country{Poland}, \orgdiv{Faculty of Physics}}
\affil[2]{\orgname{Warsaw University of Technology}, \city{Warsaw}, \country{Poland}, \orgdiv{Centre for Credible AI}}


\abstract{Information overload (IOL) is a well-known and devastating phenomenon that alters the performance of carrying out all types of tasks. It has been shown that in the media space, IOL can contribute to news fatigue and news avoidance, which often lead to the proliferation of fake news posts on social networks. However, there is a lack of automated methods for tracking IOL in large datasets. In this study, we examine whether metrics derived from the distribution of topics generated by the BERTopic model can serve as a proxy for IOL. We test our assumptions across a set of Reddit communities related to the COVID-19 pandemic and observe significant global correlation between our metrics and the FakeBERT-detected fake news fraction. However, at the community level, the correlation analysis results are ambiguous.}




\maketitle

\section{Introduction}\label{sec1}

While the effortless availability of information revolutionized our lives in recent years, another major outcome of this information revolution—realized via rapid technological advances and increased global digital connectivity — is the increasingly large volume of information encountered daily. ChatGPT, vaccines against COVID-19 and the war in Ukraine are very different examples of topics still generating masses of information from multiple sources, often non-experts. Through all sources, the individual has a constant flow of personal and professional information. Presumably, some is retained, some is used, and some is forwarded without being fully read. This ``dark side'' of information \cite{Bawden2009} entails the abundance of data beyond one’s capacity to process it, leading to the information overload (IOL) problem. This issue had troubled humanity long before the invention of print \cite{Blair2011} and has been examined from different perspectives, ranging from psychology \cite{Miller1956} to business and management sciences \cite{Roetzel2019} to journalism \cite{Briun2021}. Nevertheless, over the last 20 years, IOL has seen the penetration of social media, mainly due to rapid, ubiquitous Internet access, enabling users to send billions of comments and messages every day. IOL negatively impacts individuals, and, by cascading effects, makes social circles and whole societies vulnerable, and lately has been compared to environmental pollution \cite{IOL}. For example, a particular concern is distinguishing genuine news from fake news and manipulated messages. Social media data has been used widely in studying information overload and its consequences \cite{nematzadeh_information_2019, cinelli_covid-19_2020, gomez_rodriguez_quantifying_2014, feng_competing_2015, liang_information_2017, bermes_information_2021, tang_fake_2021}.\\

\noindent It is much easier to understand information overload intuitively than to actually define it. There are at least seven recognized definitions of information overload used in different studies \cite{eppler_concept_2004}. Probably one of the most useful is the one delivered by the cognitive load theory, suggesting that the human working memory is limited to approximately seven units of information \cite{Atkinson1968}. In this context, information overload occurs when the amount of information exceeds the working memory of the person receiving it \cite{Arnold2023,Graf2021} therefore suggesting a characteristic information capacity (budget).\\

\noindent Another stylized fact connected to IOL is the relation between the performance or decision making and the information load. It is rather intuitive to assume that the goal function is low either when one receives no information or when one is flooded with data. As a result, such a relation can take an ``inverted U shape'' \cite{Roetzel2019}, suggesting the existence of an optimal level of information load needed for best performance. This phenomenon was observed in the Usenet \cite{Jones2004} as well as Internet Relay Channels (IRC) discussions \cite{Jones2008} and lately in the case of the Twitch video-streaming platform \cite{nematzadeh_information_2019}. Analysis of Twitch detects information overload by measuring the number of messages posted by commenters. Average user's activity starts to decrease above 30 messages per second posted in the stream chat, above which information overload conditions are present \cite{nematzadeh_information_2019}. In all these cases, user activity reveals a non-monotonic behavior with respect to the number of messages present in the system.\\

\noindent A separate, important social phenomenon that haunts current media space is the spread of misinformation \cite{Zhou2020}. Dissemination of false information has a negative effect on public decision-making in public health \cite{martin_any_2022, liang_efficacy_2020}, politics \cite{allcott_social_2017}, and climate change \cite{treen_online_2020}. Especially during the COVID-19 pandemic, the spread of anti-vaccine, anti-mask, etc. narratives decreased the adherence to government mandates and recommendations, which could have had a negative impact on disease spread prevention \cite{boulos_effectiveness_2023, kafadar_determinants_2023}.\\

\noindent This study is motivated by observations linking information overload with fake news \cite{Bermes2021,Tandoc2023}. In particular, it has been suggested that IOL is connected with news fatigue that, in turn, leads to difficulties in processing the acquired information. The last issue can be associated with news avoidance \cite{Song2017,Park2019} and, in effect, increased belief in COVID-19 misinformation \cite{Bermes2021}On the other hand, it has also been observed that IOL-induced psychological strain could increase the likelihood of fake news sharing \cite{Tandoc2023}.\\

\noindent However, the mentioned studies relied on surveys conducted among the users of social media platforms in Germany \cite{Bermes2021} and Singapore \cite{Tandoc2023}. Although online questionnaires can be regarded as the most reliable source of information, as they allow for asking very specific questions, instead of relying on a proxy, they are cost- and time-consuming, and rarely repeated over time and different social media platforms.\\

\noindent Therefore, we stress the need to develop tools that measure the correlation between information overload and fake news across large datasets and can be applied instantly and online. While there are several automatic or semi-automatic approaches to account for fake news or misinformation \cite{Zhou2020} (i.e., knowledge-based, style-based, propagation-based, source-based), currently, we lack targeted methods to track information overload.\\

\noindent This study aims at (1) proposing a method to quantify information overload in large sets of online data, (2) examining a relation between IOL and fake news rate in such data. To achieve these goals, we first describe our dataset (a large sample from the Reddit platform) and the selected threads used in our analysis, and then provide details on the methods used to detect fake news based on the FakeBERT classifier. We present our original approach for quantifying information overload based on topic analysis using BERTopic and the Gini index. Finally, we present results showing the level of correlation between the IOL measure and the fake news fraction. 

\begin{table}[]
\begin{tabular}{|l|l|l|l|}
\hline
community name & number of posts & community name & number of posts\\
\hline
Coronavirus           & \num{440579}          &
COVID19positive       & \num{93576}           \\
CoronavirusCirclejerk & \num{69366}           &
CoronavirusDownunder  & \num{60543}           \\
CoronavirusUS         & \num{59289}           &
COVID19               & \num{52005}           \\
CoronavirusUK         & \num{51127}           &
covidlonghaulers      & \num{47434}           \\
CoronavirusMemes      & \num{46314}           &
CovIdiots             & \num{38076}           \\
CovidVaccinated       & \num{30648}           &
ChurchOfCOVID         & \num{28377}           \\
CanadaCoronavirus     & \num{26164}           &
COVID19\_support      & \num{17388}           \\
COVID                 & \num{15908}           &
CoronavirusCA         & \num{13108}           \\
FloridaCoronavirus    & \num{12577}           &
CoronavirusWA         & \num{12102}           \\
CoronavirusMa         & \num{11152}           &
CoronaVirusTX         & \num{10599}           \\
CoronavirusRecession  & \num{9662}            &
CoronavirusFOS        & \num{9425}            \\
CoronavirusMichigan   & \num{9256}            &
Coronavirus\_NZ       & \num{8831}            \\
CoronavirusColorado   & \num{8173}            &
CoronavirusGA         & \num{7944}            \\
nycCoronavirus        & \num{7886}            &
Coronaviruslouisiana  & \num{7598}            \\
CoronavirusCanada     & \num{7497}            &
Coronavirus\_Ireland  & \num{6699}            \\
CoronavirusIllinois   & \num{6394}            &
CoronavirusAZ         & \num{6281}            \\
LongCovid             & \num{5933}            &
CoronavirusAustralia  & \num{5020}            \\
CoronaVirusPA         & \num{3805}            &
CACovidRentRelief     & \num{3459}            \\
CoronavirusMN         & \num{3415}            &
Covid19\_Ohio         & \num{3107}            \\
COVID19PGH            & \num{2810}            &
COVIDAteMyFace        & \num{2425}\\
\hline
\end{tabular}
\caption{Names of communities (subreddits) selected for our dataset together with their sizes (number of posts).}\label{tab:subreddits}
\end{table}
\section{Datasets}
\subsection{Pushshift Reddit Dataset} Reddit is a forum social media platform with more than 110 million daily active unique users. The platform is organized into so-called \textit{subreddits}, which can be seen as separate communities or thematic fora, where registered users can create posts, sparking discussions as other users can submit comments on those posts. In 2023, there has been a widespread change in application programming interface (API) access for third parties on Twitter (X), Reddit, and TikTok; as a consequence, a lot of social media data, until now taken for granted, became inaccessible to researchers. Those changes made it hard for researchers to engage with Open Science (OS) practices \cite{davidson_platform-controlled_2023}. Because of those changes and data engineering challenges associated with API use, Pushshift Reddit Dataset \cite{baumgartner_pushshift_2020} was created, originally consisting of 20~000 top subreddits spanning between June 2006 and December 2022 \cite{PRD2025} (a new version consists of 40~000 subreddits \cite{PRD2025new}). Ever since its publication, the dataset has been used in a wide variety of research, such as generating text embeddings for characterization of online communities \cite{Sawicki2023}, spread of misinformation in online communities during COVID-19 pandemic \cite{cinelli_covid-19_2020}, topic modeling of user content \cite{kedzierska_topic_2024}, mental illness discourse \cite{de_choudhury_mental_2014}, temporary identity and anonymity perceptions \cite{leavitt_this_2015}. Reddit has been proven to be a quality source of information in almost all fields of interest \cite{sawicki_exploring_2022}.

\subsection{Reddit COVID-19 Dataset} We selected Reddit communities (subredits), with keywords \enquote{covid} and \enquote{coronavirus} in their names out of the Pushshift Reddit Dataset. As a result, \num{40} subreddits with \num{1261952} posts were downloaded. Community sizes vary greatly from small, like \enquote{COVIDAteMyFace} \num{2425} posts, to large, like \enquote{Coronavirus} \num{440579} posts (see Table \ref{tab:subreddits}). Post activity also varies greatly in time, with large amounts of content being shared in the first year of the pandemic (see Results section for details).

\subsection{COVID-19 Rumor Dataset} COVID-19 Rumor Dataset \cite{Rumor2025} is a manually labeled collection of rumors \num{6834} data from news websites and tweets. Data labeling was based on information accessible on fact-checking websites like \href{poynter.org}{poynter.org} or \href{factcheck.org}{factcheck.org}. Based on this dataset, Cheng et al. have developed a machine learning algorithm for veracity, stance and sentiment classification \cite{cheng_covid-19_2021}. Similar text style is present on both Reddit and Twitter, as well as comparable length, making it possible to use this dataset for training a Reddit classifier. The dataset has been used in subsequent studies connected to COVID-19 fake news and rumors \cite{Mahbub2022,Kochkina2023,Timoneda2025}.

\section{Measures}

\subsection{Information overload} As mentioned in the Introduction section, there is a lack of general measures connected to information overload, which additionally depends on the targeted situation. For example, in the case of science, researchers acknowledge that an exponentially growing number of papers \cite{Fortunato2018} results in the overproduction of information \cite{Farber2023}, popularly nicknamed as ``paper tsunami'' \cite{Brainard2020}. In the case of the examined system, it would be equivalent to the total number of posts $PC$ in a given unit of time (e.g., one day or one week). Another option is to consider the number of different topics $TC$ discussed at a given time or the ratio of topics to posts $TC/PC$ (with $1/PC \le TC/PC \le 1$). Previous works strongly suggest the ``inverted U shape'' \cite{Roetzel2019} observed in the case of Twitch video-streaming platform \cite{nematzadeh_information_2019}, however, in such a case one needs to propose a specific shape of the performance function and additionally motivate its parameters based on some observations, e.g., by a survey or other information.\\

Neither of the discussed indicators considers the distribution of the topics and their diversity. To account for these characteristics, we have decided to use the Gini coefficient (Gini index), a common measure of wealth inequality \cite{Ceriani2012}, as our proxy for the information overload indicator. Assuming that at a given moment there are $TC$ topics discussed using $PC$ posts and $x_{i}$ denotes the number of posts in the $i$-th topic and that $x_{i}$ is arranged according to increasing number of posts, we can use the following definition \cite{Damgaard2000}

\begin{equation}
    G = \frac{\sum\limits_{i=1}^{TC}(2i - TC - 1)x_{i}}{TC \cdot PC}
\label{eq:gini}
\end{equation}
where, $PC = \sum_{i=1}^{i=TC}x_{i}$. Small Gini coefficient values indicate that the discussed topics are of more similar sizes. On the other hand, the Gini index reaching 1 points to the fact that although there is a specific number of discussed topics, one of them monopolizes the discussion space. The metric has been calculated for each week, and over a whole time frame present in the dataset. To calculate the number of topics in the analyzed posts of our dataset we used BERTopic \cite{grootendorst_bertopic_2022} algorithm (see "Models and algorithms: Topic modeling"  for details).

Entropy is another commonly used measure for detecting imbalance in machine learning approaches (e.g., diversity measure for CARTs \cite{Hastie2001}) or in more general cases (e.g., indicator of the phase of the dialogue based on probabilities of specific emotional valences \cite{Sienkiewicz2013}). In our case, assuming that $p_i$ denotes the fraction of posts in the $i$-th topic, entropy can be expressed as

\begin{equation}
    H = - \sum\limits_{i=1}^{i=TC} p_i \log p_i
\label{eq:entropy}
\end{equation}

\begin{figure}[h]
    \centering
    \includegraphics[width=0.75\linewidth]{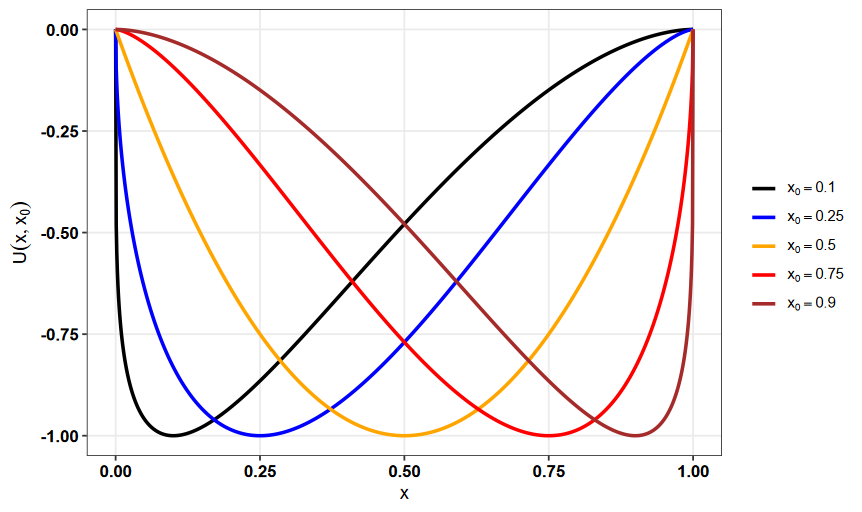}
    \caption{$U(x,x_0)$ function for different values of the $x_0$ parameter.}
    \label{fig:fig1}
\end{figure}

In line with the previous description, we have also taken into account the above-mentioned ``inverted U shape'' characteristic of the decision-making performance function in relation to information load. However, we have decided to introduce the following changes: (1) we invert this measure to reflect information overload rather than the performance, (2) as a dependent variable we use the ratio of topic to posts $x=TC/PC$, (3) we parameterized the function to take into account the fact, that the optimum does not necessarily need to appear for $x=1/2$. Taking into account these, we propose the following function $U(x,x_0)$

\begin{equation}
    U(x,x_0) = -\left(\frac{x}{x_0}\right)^{2x_0}\left(\frac{1-x}{1-x_0}\right)^{2(1-x_0)}
\label{eq:u}
\end{equation}
shown for various values of the $x_0$ parameter in Fig.~\ref{fig:fig1}. Let us note that for $x_0=1/2$, Eq. (\ref{eq:u}) reduces to a parabola $U(x)=-4x(1-x)$.

\subsection{Fake news} To account for the veracity of the posts in a given subreddit or in the whole dataset for each week, we calculated the fraction of fake news $f$ using the FakeBERT \cite{kaliyar_fakebert_2021} classifier (see Sec. \ref{sec:models_fake} for details).

\section{Models and algorithms}
Bidirectional encoder representations from transformers (BERT) \cite{devlin_bert_2018} language model has recently been used in a variety of natural language processing (NLP) tasks. In a similar way as Generative Pre-trained Transformer (GPT) models, BERT excels in next sentence prediction. During the pre-training phase on large datasets, BERT learns contextual, latent representations of tokens in their context. After the pre-training BERT can be fine-tuned with smaller datasets to perform such actions like named entity recognition \cite{sawicki_applying_2024}, topic modeling \cite{grootendorst_bertopic_2022}, text classification \cite{garrido-merchan_comparing_2023}, sentiment analysis \cite{alaparthi_bert_2021}, machine translation \cite{zhu_incorporating_2020} and more.\\ 

\noindent In this study we make use the BERT model to automatically detect topics in our selected dataset as well as to classify the fake news content.   

\subsection{Topic modeling}\label{sec:models_topic} A topic model is an algorithm that allows for  an automatic (unsupervised) detection of concepts (named as topics) that occur in several documents. For purposes of topic detection in our study, BERTopic \cite{grootendorst_bertopic_2022} modeling technique was selected. BERT enhances class-based TF-IDF by using document embeddings and BERT embeddings have a large advantage over bag-of-words approach by including the semantic relationships among words.\\

\begin{table}
    \centering
    \begin{tabular}{|l|r|r|r|r|}
    \hline
    class & precision & recall & $F_1$ score & support \\
    \hline
    F     & 0.8551    & 0.8969 & 0.8755   & 737     \\
    U     & 0.6166    & 0.6204 & 0.6185   & 324     \\
    T     & 0.6795    & 0.6107 & 0.6433   & 375\\
    \hline
    \end{tabular}
    \caption{Evaluation metrics of FakeBERT model on the COVID-19 Rumor Dataset}
    \label{tab:classification_report}
\end{table}

\noindent Reddit's structure provides a natural way to divide the datasets. Content on the websites is divided into \enquote{subreddits}, henceforth referenced as, communities. This natural differentiation of topics provided by the site's user base might provide valuable nuance that might have been lost while applying BERTopic to the whole dataset at ounce. Additionally, using a full dataset model requires as much as 8 times more memory (RAM), and takes as much as 13 times longer to process. Full dataset models are also too large to run on consumer GPUs and require specialized hardware. As a consequence, two modeling techniques were tested, per community modeling referenced as \enquote{Ds} and full dataset modeling techniques referenced as \enquote{F*} (see Fig. \ref{fig:fig2}c-d). \enquote{Fd} techniques use distribution strategy for outlier reduction and \enquote{Fe} use embeddings, both of which with similar results. An inseparable part of any internet forum is moderation, and Reddit isn't any different in that respect. Posts on the website cannot be outright removed, because that would break the quote post functionality. As an alternative to outright deletion, Reddit staff replaces contents breaking the site's policy with \enquote{[user removed]} and if the user chooses to remove a post by themselves it is replaced with \enquote{[user deleted]}. This raises a dilemma, because of the nature of the dataset, removed content has high probability of having been harmful or fake, but it can as well be something irrelevant like spam or unwanted bot activity. Should those posts be included in the analysis or discarded? Both possibilities have been explored, models with \enquote{r} in their name discard those posts from their corpus, while the rest include it.

\subsection{Fake news detection}\label{sec:models_fake} 
\begin{figure}
    \centering
    \includegraphics[width=.9\textwidth]{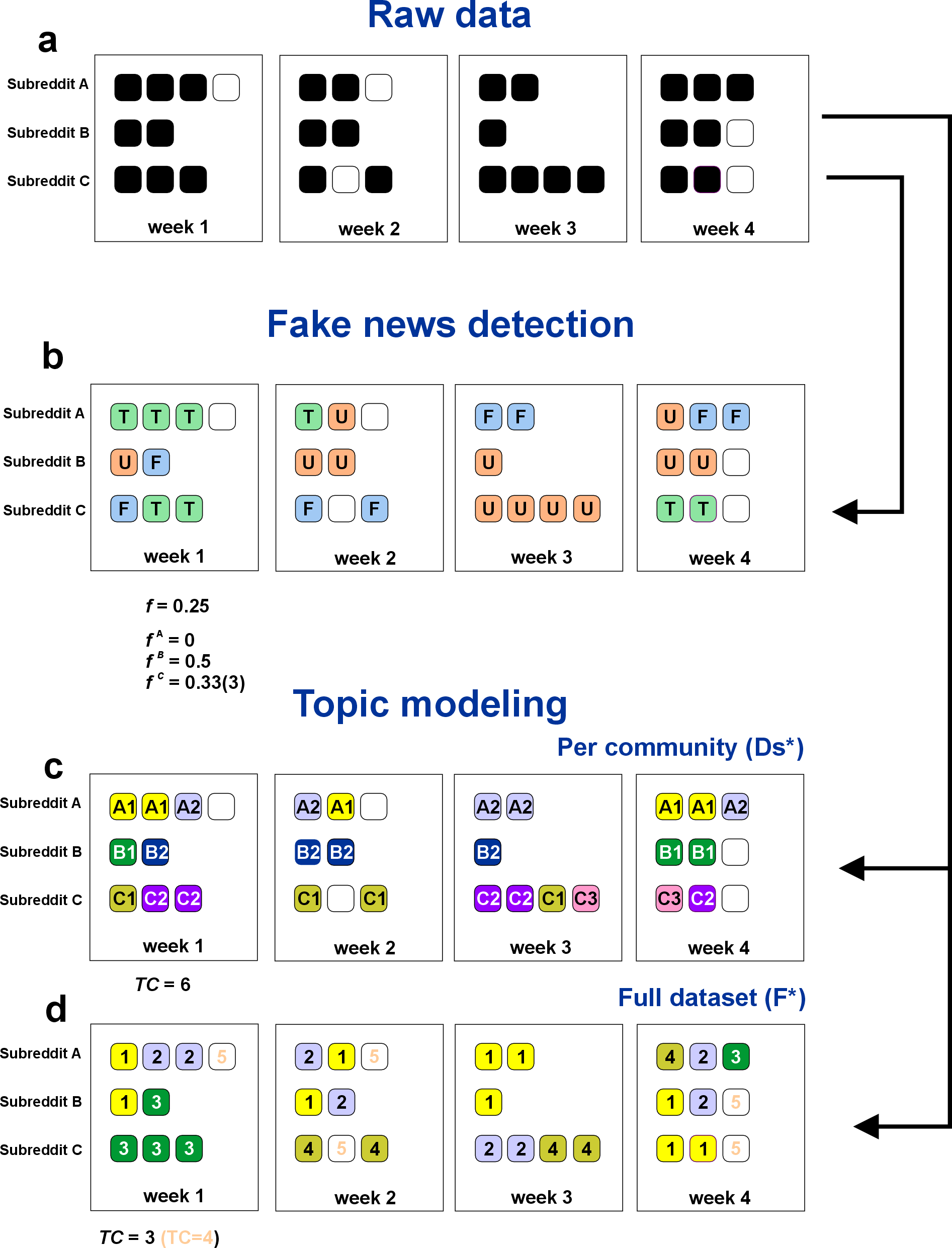}
    \caption{The pipeline of the data processing. (a) Raw data comes as posts in different subreddits published in a given week. Full symbols represent normal posts, empty ones denote posts with replaced content (user removed / user deleted). (b) Using the FakeBERT model we assign labels \textit{true}, \textit{fake} and \textit{unverified} to all posts with unaltered content. Based on that we can calculate the global fraction of fake posts $f$ in a given week or do it at the community level ($f^A$ etc). (c-d) We use the BERTopic algorithm to detect topics in the posts in an unsupervised manner. Versions labeled $Ds^*$ (panel c) work at the community level, i.e., the topic detection algorithm is run on each subreddit separately, while $F^*$ versions run on the whole dataset (panel d). Additionally models denoted with $r$ in their name discard altered content while others include them.}
    \label{fig:fig2}
\end{figure}

BERT classification algorithm used in this study is based on the FakeBERT method \cite{kaliyar_fakebert_2021}. FakeBERT architecture utilizes BERT embeddings as the first step, then running them through a classifier head consisting of convolution and max pooling layers, with fully two connected layers ending with the last fully connected layer of output size three (fake, true, unverified classes). As a result, each document is assigned a truthfulness class (see Fig. \ref{fig:fig2}b).\\

\noindent Veracity classification results, reported in the COVID-19 Rumor Dataset study, have been reproduced. The classifier works best at finding fake information, while having more trouble discerning true and unverified information. Fake class carries the most importance in this research; therefore, classification results can be applied to further analyses. Harmonic mean of precision and recall, i.e., $F_1$ scores for each class, have been represented in Table \ref{tab:classification_report}. The classifier was trained using a k-fold cross-validation method; for 10 epochs, just the classifier head was trained, and 5 epochs of fine-tuning of the whole neural network.

\begin{figure}[!ht]
\includegraphics[width=\textwidth]{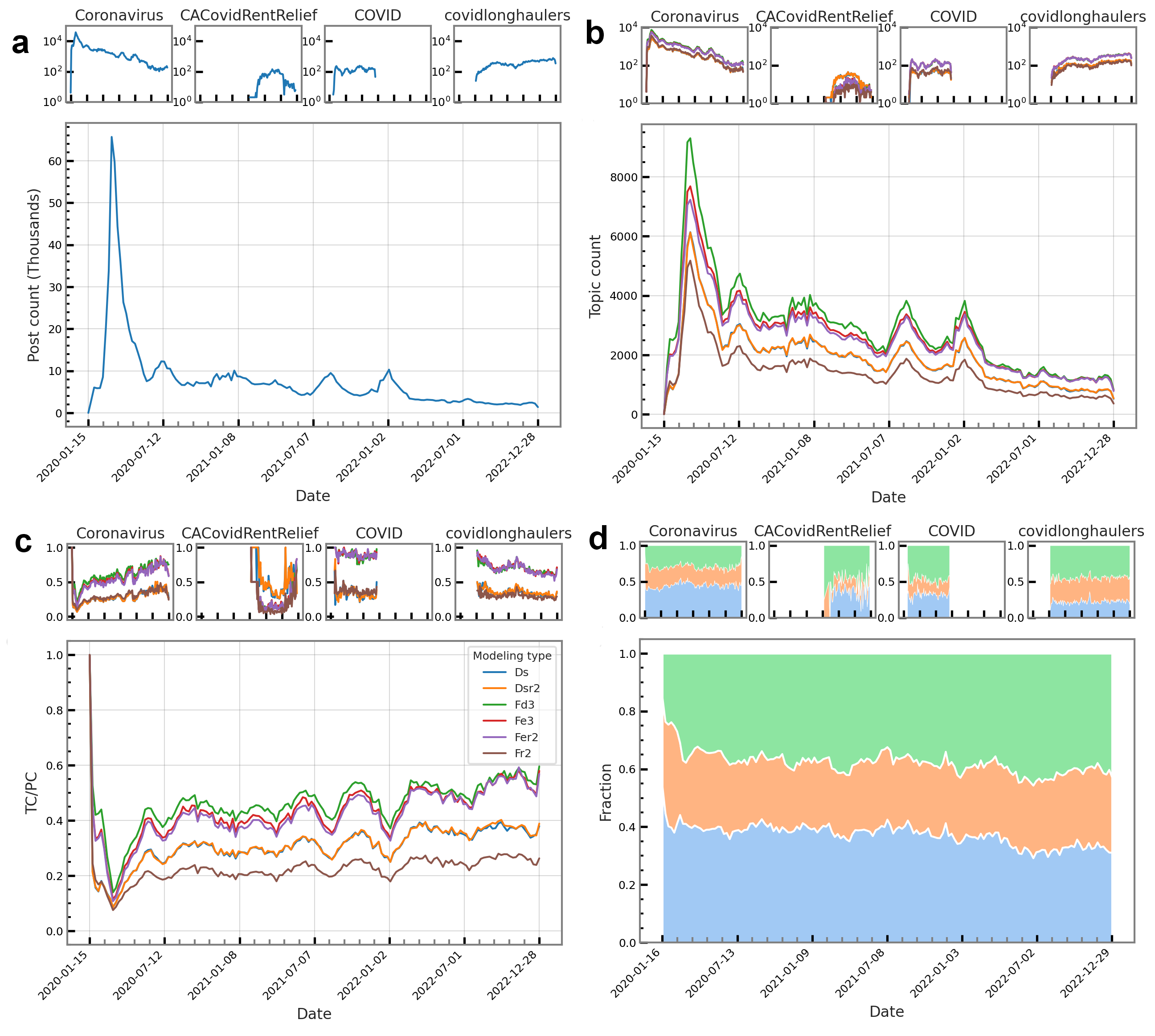}
    \caption{Fundamental metrics of the dataset across time: (a) The total number of posts published in each subreddit (b)  The total number of topics generated using different modeling techniques. (c) The ratio of number of topics to the number of posts $TC/PC$. (d) The fraction of the true posts (green), fake posts (blue) and unverified ones (orange). All statistics aggregated in weekly intervals. Panels (b-d) show results for different parametrization of the BERTopic model (see Sec. \ref{sec:models_topic} for details). Four selected communities have been represented in the insets.}
    \label{fig:fig3}
\end{figure}

\section{Results}

\subsection{Post and topics}
Let us start by once again bringing to the front a great variability of the data as well as its strong diversity. Around half the overall posts published come from the first year of the pandemic (2020), with a substantial drop off in interest over time, as presented in Fig. \ref{fig:fig3}a, where we show the total number of posts per week together with insets giving an overview of four selected subreddits (chosen to represent diverse patterns in the dataset). The \textit{Coronavirus} is the largest community by a large margin -- the very large influx of posts in the first months of the pandemic created a visible peak of around 40,000 posts per week. Some communities are still growing (\textit{covidlonghaulers}) while activity on others has largely died down (\textit{CACovidRentRelief}), the others -- like \textit{COVID} became private (closed) ones.\\

\noindent In the same way, the number of topics obtained from the BERTopic algorithms varies greatly over time and across communities, as shown in Fig. \ref{fig:fig3}b. There is a substantial difference in the number of topics generated by Ds models than F* models, although the overall characteristics stay the same. Both panels also suggest a significant correlation between the number of posts and the number of topics returned by BERTopic, which can be seen as an obstacle to this study. In spite of that, as it is clearly pictured in Fig. \ref{fig:fig3}c the ratio of topics to posts $TC/PC$ varies over time, e.g., for the \textit{Coronavirus} community, we observe a steady increase during the whole pandemic period, while \textit{covidlonghaulers} presents an opposite behavior; \textit{CACovidRentRelief} does not show any of these patterns. These examples suggest that the number of topics can yield valuable information beyond the number of posts.

\subsection{Fake news} Figure \ref{fig:fig3}d presents the results of applying the FakeBERT classifier to the COVID-19 Reddit dataset. The amount of fake information seems to vary with time -- let us note that at the beginning of the pandemic only around 25\% are classified as true. During the next months, the situation gradually changes, bringing less fake news, and by the end of the examined period, true posts become the majority. As can be seen from the insets in Fig. \ref{fig:fig3}d different communities can be characterized with varied overall fake information detection rate -- while for \textit{COVID} and \textit{covidlonghaulers} the fraction of true posts oscillates around 0.5, the \textit{Coronavirus} subreddit is shifted toward fake news.

\subsection{Information overload} Although useful, the number of topics and the ratio of topics to posts might shadow the fact that the whole discussion in a given community or over the whole observed dataset could be concentrated on a single topic with a very small addition of other subjects. To account for such a behavior, we used the Gini index $G$ defined by Eq. (\ref{eq:gini}). Indeed, let us rewrite (\ref{eq:gini}) as 

\begin{equation}
    G = 2\frac{\sum\limits_{i=1}^{TC}ix_{i}}{TC \cdot PC} - \frac{TC+1}{TC} \approx 2\frac{\sum\limits_{i=1}^{TC}ix_{i}}{TC \cdot PC} - 1
\label{eq:gini1}
\end{equation}
and assume that all the topics but one are represented by a single post (i.e., $x_{i} = 1$ for $i=1,...,TC-1$), while the dominating topic gathers all the remaining posts (i.e., $x_{i} = PC - (TC-1)$ for $i=TC$). Making use of the sum of an arithmetic progression, we instantly arrive at
\begin{equation}
    G \approx 1 - \frac{TC}{PC}
\label{eq:gini2}
\end{equation}
which holds for $PC \gg 1$ and $TC \gg 1$ (and a very uneven distribution of posts to topics). Now, let us compare two situations, both characterized by 100 posts and 50 topics. In the first case, one topic gathers 51 posts, and the remaining 49 topics are represented by just one post each. Then, applying Eq. (\ref{eq:gini2}) we get $G \approx 0.5$. In the second situation, the posts are evenly spread, i.e., each topic consists of 2 posts (in general, each topic gets $PC/TC$ posts). Then, using Eq. (\ref{eq:gini1}), we have $G=0$, which is an expected result coming out of the definition of the Gini index. As we can see, although in both cases we have the same ratio $TC/PC=1/2$, Gini index makes it possible to distinguish these two extremely different situations (let us underline here that when have a very small number of topics and large number of posts not evenly spread posts, e.g., 1000 posts, 2 topics where one gets 999 posts and the other just one, the Gini index $G$ approaches 1).\\

\begin{figure}
    \centering
    \includegraphics[width=\textwidth]{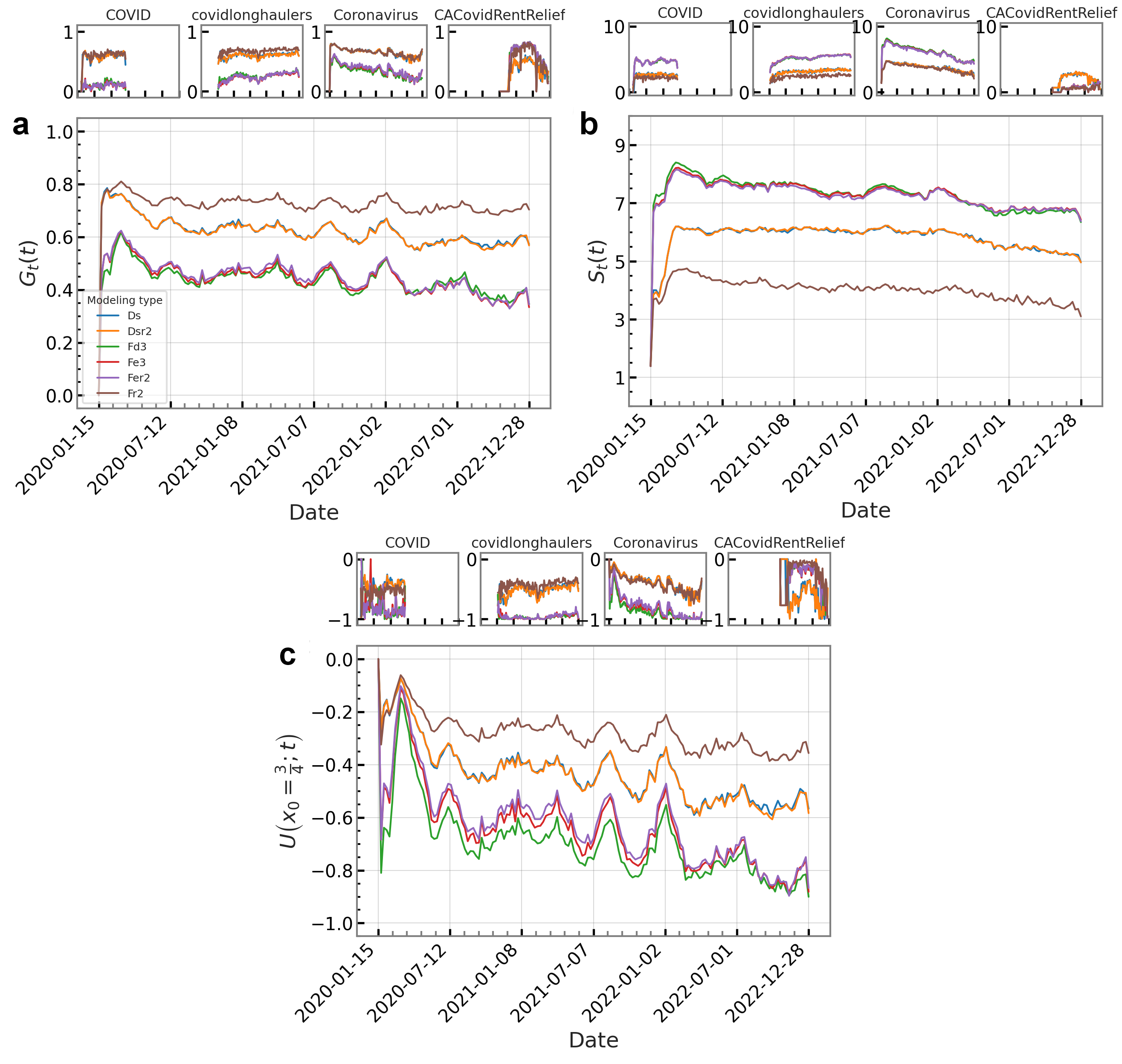}
    \caption{Information overload metrics over time: Gini index (a), entropy (b) and $U$ metric (c). Line colors reflect different versions of the BERTopic model (see Sec. \ref{sec:models_topic} for details). Four selected communities from Fig. \ref{fig:fig3} have been shown in the insets.}
    \label{fig:fig4}
\end{figure}

\noindent The results for the Gini index are presented in Fig. \ref{fig:fig4}a. Interestingly, $G$ starts from rather large values at the beginning of the pandemic (even as large as 0.8 for $Fr$, $Ds$, and $Dsr$ models) and decreases with time, while this decay is negatively correlated with the increase of $TC/PC$ ratio shown in Fig. \ref{fig:fig3}c. The last fact suggests that the system operates in the regime covered by Eq. (\ref{eq:gini2}) and hints at high information overload induced by single topics dominating the discussion space.\\

\noindent Let us now move to the results obtained using the entropy defined by Eq.~(\ref{eq:entropy}) to address information overload. Let us again compare two situations, the same as in the case of the Gini index. In the first case, with one topic gathering 51 posts, while the remaining 49 topics gathered each a single post results in $H\approx2.60$. In the second case, with equally popular 50 topics, $H$ reaches approximately $3.91$. As we can see, entropy gives similar results to the Gini index, which is confirmed in Fig. \ref{fig:fig4}b for our data, in general, regardless of the utilized topic model method.\\

\noindent Finally, the $U$ measure (for $x_0 = 3/4$) versus time is shown in Fig. \ref{fig:fig4}c, and, in general, repeats the previously mentioned time-series characteristics. All three methods suggest that  -- at least at the global level -- the Reddit community commenting on COVID-19 moves from the region of high information overload towards lower values. 

\subsection{Correlations between information overload and fake news}

Having described both information overload metrics as well as the fake information indicator, let us now assess the level of correlation between these two. To quantify it we use sample Pearson correlation coefficient, defined for a given community $i$ is 
\begin{equation}
    \rho^{i}(f, X) = \frac{\sum\limits_{t=1}^{T_i}\left(f_t-\bar{f}\right)\left(X_t-\bar{X}\right)}{\sqrt{\sum\limits_{t=1}^{T_i}\left(f_t-\bar{f}\right)^2\sum\limits_{t=1}^{T_i}\left(X_t-\bar{X}\right)^2}}
\end{equation}
where index $t$ goes over the weeks and $X$ represents one of the three metrics ($G$, $H$ or $U$) the $T_i$ denotes the number of weeks in the time series and $\bar{f}=\frac{1}{T_i}\sum_{t=1}^{T_i}f_t$, $\bar{X}=\frac{1}{T_i}\sum_{t=1}^{T_i}X_t$. The coefficient is the calculated by creating two time series for each community, where the first one contains Gini index for consecutive weeks $G_t$, while the second one the fraction of fake news $f_t$, also at weekly level. The crucial distinction comes from the way these time series are obtained.

\begin{figure}[ht]
    \centering
    \includegraphics[width=\textwidth]{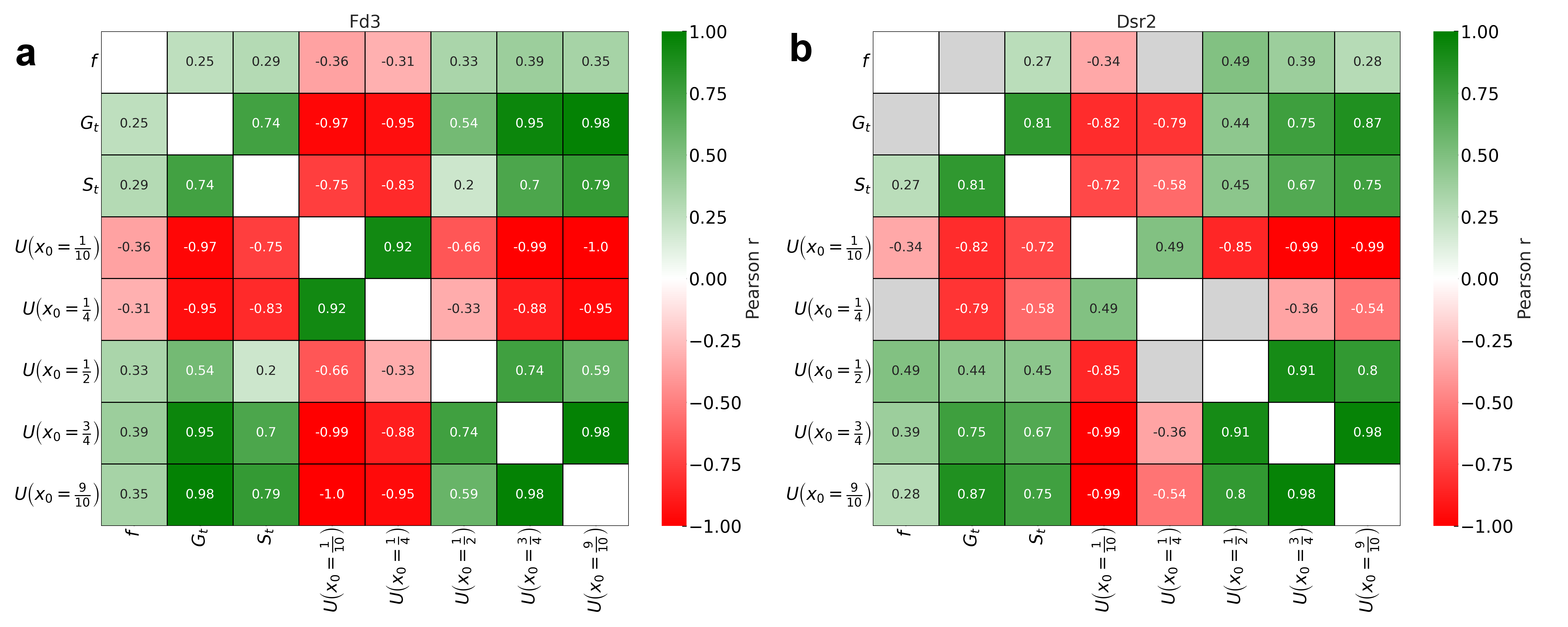}
    \caption{Global correlation coefficient matrix between the principal measures in this study: fake news fraction $f$, Gini index $G$, entropy $H$ as well as the $U$ metric for different values of $x_0$ for (a) ``Fd3'' method (full dataset and 'distributions' outlier reduction strategy) and (b) ``Dsr2'' method (subreddit level with the removal of 'user deleted' and 'user removed' phrases). Gray color reflects correlation coefficient with p-value less than 0.05.}
    \label{fig:corr}
\end{figure}

It is possible to measure the relationship between information overload and the fraction of fake news at the global level by calculating the overall correlation coefficient between these two time series. These results are shown for different metrics in Fig. \ref{fig:corr}, strongly suggesting a small role played by the type of BERT model. Indeed, regardless of the chosen method (here, full-dataset $Fd3$ and community-level $Dsr2$), global correlation values stay at a similar level, indicating a connection between fake news levels and information overload, in particular for the entropy as well as for the $U$ metrics with $x_0 \ge 1/2$.

\begin{figure}[!ht]
    \centering
    \includegraphics[width=.85\textwidth]{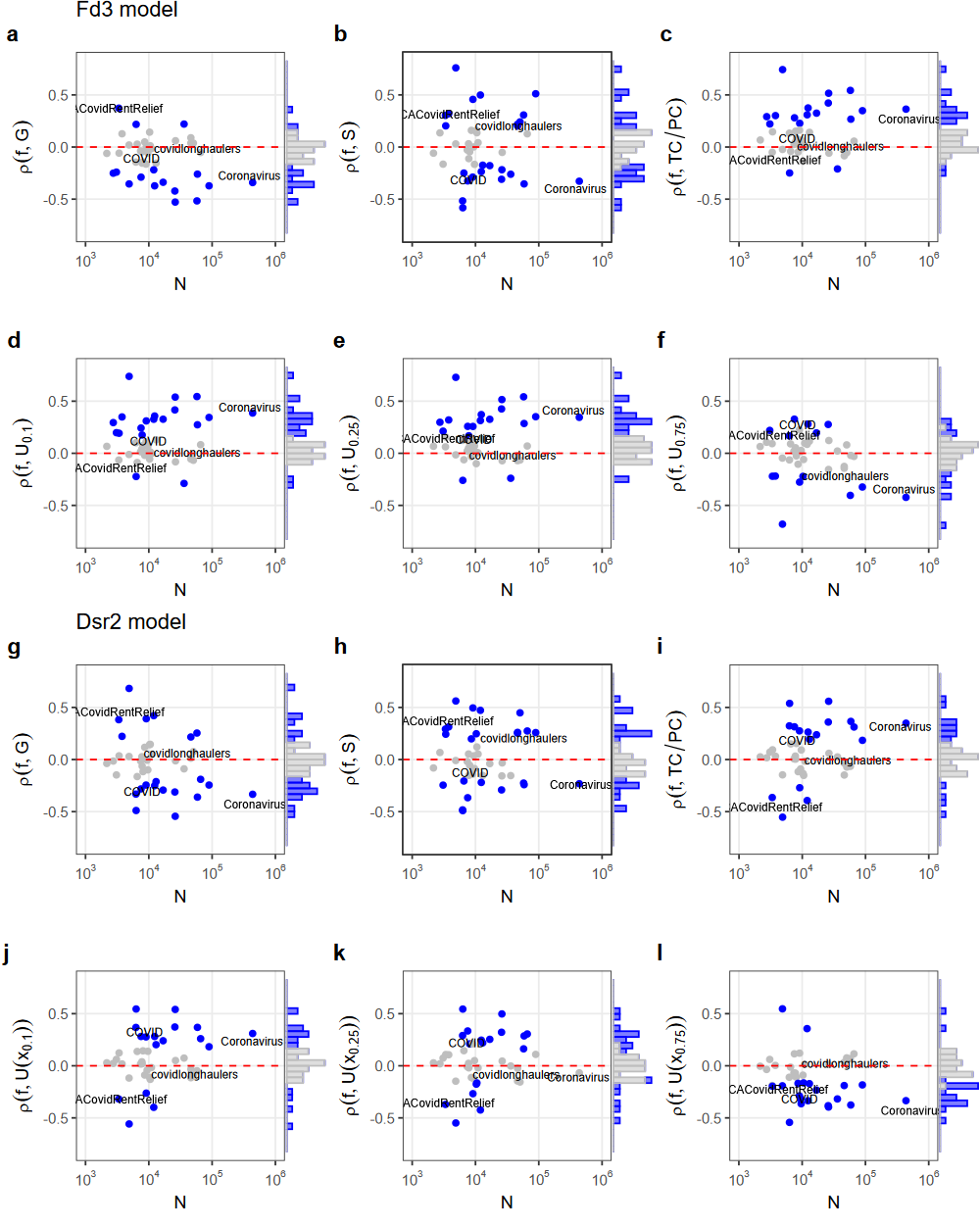}
    \caption{Pearson correlation coefficient between the fraction of fake posts at the subreddit level $f$ and selected measures of information overload versus the size of the community $N$ (total number of posts in the subreddit): Gini coefficient (panels \textbf{a} and \textbf{g}), Shannon entropy (\textbf{b} and \textbf{h}), ratio of topics to posts (\textbf{c} and \textbf{i}) and and $U$ measure for $x_0=0.1$ (\textbf{d} and \textbf{j}), $x_0=0.25$ (\textbf{e} and \textbf{k}) and $x_0=0.75$ (\textbf{f} and \textbf{l}). Each data point represents one subreddit, the four subreddits mentioned in Figs. \ref{fig:fig3}-\ref{fig:fig4} are marked. The gray color represents correlation coefficients with p-value $<.05$. The two top rows (\textbf{a-f}) are the $Fd3$ method, where the topic distribution was generated on the whole dataset; the two bottom rows (\textbf{g-l}) is the \texttt{Dsr2} method, with the topic distribution generated per community. The dashed line marks $\rho=0$; the right y-axis shows the histogram of correlation coefficients.}
    \label{fig:fig6}
\end{figure}

\noindent This picture changes dramatically when we move from global time series to local ones. Detailed results shown in Fig. \ref{fig:fig6}, where the correlation coefficient is calculated at the community level (i.e., for each subreddit separately), suggest that the ratio of topics to posts and the $U$ metric with $x_0 \le 0.5$ are strongly connected to fake news time-series. Both Gini index and entropy bring ambiguous results in contrast to the global level analysis.

\section{Conclusion and discussion}

In this paper, we have presented a comprehensive study on the Reddit communities connected to COVID-19 pandemic, covering three years between 2020 and 2022. Our results show that these communities are highly diverse, share different patterns of posting, probably linked to the subject they discuss (e.g., long-term effects of COVID, question of COVID rent relief). Moreover, on the basis of a dedicated fake news detection method, we have also shown the level of fake information in the examined subreddits.\\

\noindent The main goal of this study was to (1) test how topic modeling can be used to indicate the level of information overload (IOL) and (2) check if such an approach is sufficient to obtain the correlation between IOL and fake news reported in previous studies \cite{Bermes2021,Tandoc2023}. The algorithmic techniques based on BERT model to detect topics in Twitter posts connected to COVID-19 have been frequently used before \cite{Wang2020,Ng2022}, their primary goal was set to track the evolution of specific topics and to check how the sentiment of such topics links with, e.g., the number of infected individuals or administered vaccines. Instead, we have paid attention to the distribution of posts among topics, trying to build an information overload metric upon it.\\

\noindent Our results on the correlation between the proposed IOL metrics and the fake news fraction at the global level seem to confirm the findings of surveys \cite{Bermes2021,Tandoc2023}. However, while we move the analysis to the subreddit level, the effect is far from clear. There are several explanations for such an outcome. The first and most obvious is that, due to large differences in community size, the global analysis is heavily influenced by the largest communities, such as \textit{Coronavirus}. Secondly, Reddit posts might be a poor source for IOL/fake news detection -- it may be necessary to go to the comment level to obtain more reliable results.  A final option is to track user information in the same way as in the Twitch study \cite{nematzadeh_information_2019} or in the BBC posts analysis and modeling \cite{Chmiel2011a}, where statistics on unique users have been used.

\section*{Ethics approval and consent to participate}
Not applicable.

\section*{Consent for publication}
Not applicable.

\section*{Availability of data and material}
All the data are publicly available \cite{PRD2025}\\The modified FakeBERT model used to produce the results presented in this work is available at:\\\url{https://huggingface.co/jrawa/fake-distilbert-3class}. \noindent BERTopic implementations as well as generated topics available at:\\\url{https://gitlab.com/jan_rawa/reddit-topic-modeling}.

\section*{Competing interests}
The authors declare no competing interests.

\section*{Funding}
\textbf{J.R.} and \textbf{J.S.} acknowledge support by POB Cybersecurity and Data Science of Warsaw University of Technology within the Excellence Initiative: Research University (IDUB) programme. Addionally \textbf{J.S.} acknowledges support by the European Union under the Horizon Europe grant OMINO – Overcoming Multilevel INformation Overload (grant number 101086321, \href{https://ominoproject.eu}{https://ominoproject.eu}). Views and opinions expressed are those of the authors alone and do not necessarily reflect those of the European Union or the European Research Executive Agency. Neither the European Union nor the European Research Executive Agency can be held responsible for them. The computational part of this study was supported in part by the Poznań Supercomputing and Networking Center.

\section*{Authors' contributions}
\textbf{J.R.} and \textbf{J.S.} conceived and planned the study. \textbf{J.R.} wrote the draft version of the manuscript, performed experiments, and wrote the code. All authors reviewed the manuscript.

\bibliography{sn-bibliography}

\end{document}